\documentclass{article}

\usepackage{arxiv}
\usepackage{listings}
\usepackage{xcolor}
\usepackage[utf8]{inputenc} 
\usepackage[T1]{fontenc}    
\usepackage{url}            
\usepackage{booktabs}       
\usepackage{amsfonts}       
\usepackage{nicefrac}       
\usepackage{microtype}      
\usepackage{lipsum}
\usepackage{multicol}
\usepackage{multirow}
\usepackage{capt-of}
\usepackage{enumitem}
\usepackage{graphicx}
\usepackage[draft]{minted}
\usepackage{courier}
\usepackage{hyperref}       

\definecolor{QM_grey}{RGB}{225,225,225}
\definecolor{QM_blue}{RGB}{48,130,207}
\definecolor{darkgray}{rgb}{.4,.4,.4}
\definecolor{QM_green}{RGB}{0,130,60}
\definecolor{QM_Lgreen}{RGB}{191,250,221}
\definecolor{QM_orange}{RGB}{245,132,31}
\definecolor{QM_red}{RGB}{240,0,60}
\definecolor{QM_purple}{RGB}{121,19,229}
\definecolor{backcolour}{rgb}{0.95,0.95,0.92}

\lstdefinestyle{mystyle}{
    backgroundcolor=\color{QM_grey},   
    commentstyle=\color{QM_green},
    keywordstyle=\color{magenta},
    numberstyle=\tiny\color{QM_blue}, 
    stringstyle=\color{codepurple},
    basicstyle=\ttfamily\footnotesize,
    breakatwhitespace=false,         
    breaklines=true,                 
    captionpos=b,                    
    keepspaces=true,                 
    numbers=left,                    
    numbersep=2pt,                  
    showspaces=false,                
    showstringspaces=false,
    showtabs=false,                  
    tabsize=2
}

\lstdefinelanguage{python}{
  keywords={n, tau, x, discr_threshold, state, states, qubit_freq, freq_correction, s, feedback_latency, ce_time, re_time, sampling_window, time_of_flight, frame_rot_ang, frame_lut, i, H, f, H, H0, T, s, s_ar, frame_rot_ang_ar, frequency_ar, frequency_lut, frequency_update, amp_ar, amp_lut, amp_update, x1, x2, threshold_ar, threshold_lut, states_vec},
  keywordstyle=\color{QM_blue}\bfseries,
  keywords=[2]{readout_pulse, resonator, pi, pi_half, qubit, readout_element, control_element, control_pulse, all_elements, readout_elements, control_elements},
  keywordstyle=[2]\color{QM_green}\bfseries,
  keywords=[3]{measure, wait, play, align, frame_rotation_2pi},
  keywordstyle=[3]\color{QM_red}\bfseries,
  keywords=[4]{tau_max, n_avg, max_latency, N_inout, N_in, N_out, N_shots},
  keywordstyle=[4]\color{QM_orange}\bfseries,
  keywords=[5]{for, if, while, in},
  keywordstyle=[5]\color{QM_purple}\bfseries,
  commentstyle=\color{darkgray}\ttfamily,
}

\hypersetup{
    colorlinks=true,
    linkcolor=QM_red,
    filecolor=magenta,
    citecolor=QM_blue,
    urlcolor=QM_red,
    pdftitle={Overleaf Example},
    pdfpagemode=FullScreen,
    }

\makeatletter
\def\bstctlcite{\@ifnextchar[{\@bstctlcite}{\@bstctlcite[@auxout]}}
\def\@bstctlcite[#1]#2{\@bsphack
  \@for\@citeb:=#2\do{%
    \edef\@citeb{\expandafter\@firstofone\@citeb}%
    \if@filesw\immediate\write\csname #1\endcsname{\string\citation{\@citeb}}\fi}%
  \@esphack}
\makeatother

\title{Quantum-classical processing and benchmarking\\ at the pulse-level}

\author{
 Lior Ella, Lorenzo Leandro, Oded Wertheim, Yoav Romach, Lukas Schlipf,\\
 \textbf{Ramon Szmuk, Yoel Knol, Nissim Ofek, Itamar Sivan and Yonatan Cohen\thanks{Corresponding author: yonatan@quantum-machines.co} }\\
  \\
  \textit{Quantum Machines Inc., Tel Aviv, Israel}\\
  \\
}

\begin{document}
\bstctlcite{IEEEexample:BSTcontrol}
\maketitle

\begin{abstract}
Towards the practical use of quantum computers in the NISQ era, as well as the realization of fault-tolerant quantum computers that utilize quantum error correction codes, pressing needs have emerged for the control hardware and software platforms. In particular, a clear demand has arisen for platforms that allow classical processing to be integrated with quantum processing. While recent works discuss the requirements for such quantum-classical processing integration that is formulated at the gate-level, pulse-level discussions are lacking and are critically important. Moreover, defining concrete performance benchmarks for the control system at the pulse-level is key to the necessary quantum-classical integration. In this work, we categorize the requirements for quantum-classical processing at the pulse-level, demonstrate these requirements with a variety of use cases, including recently published works, and propose well-defined performance benchmarks for quantum control systems. We utilize a comprehensive pulse-level language that allows embedding universal classical processing in the quantum program and hence allows for a general formulation of benchmarks. We expect the metrics defined in this work to form a solid basis to continue to push the boundaries of quantum computing via control systems, bridging the gap between low-level and application-level implementations with relevant metrics. 
\end{abstract}
\vspace{1cm}

\begin{multicols}{2}

\section{Introduction} \label{CH1: Intro}
Integration of classical processing has been a requirement for quantum computing since its inception. This ever-present need for quantum-classical integration spans a gamut from the fundamental to the practical. For example, it allows for more efficient preparation of states with long range order\cite{Piroli2021, Hoyer2005} and for routing\cite{Devulapalli2022} using constant depth circuits, than is possible using local unitary operations alone\cite{Piroli2021, Bravyi2006}. It is also an essential requirement of measurement-based quantum computation\cite{Raussendorf2001} as well as of quantum error correction (QEC) codes\cite{Ofek2016} and fault tolerance\cite{Ryan2021}. Hybrid quantum-classical variational algorithms, such as variational quantum eigen-solvers (VQE) and quantum approximate optimization algorithms (QAOA), are an integral part of noisy intermediate scale quantum (NISQ) computation\cite{Bharti2022, Cerezo2021, Wack2021}. Additionally, integrating classical processing allows performing algorithms such as phase estimation with increased efficiency\cite{Corcoles2021, Granade2022, Lubinski2022}, resolving Fock states in superconducting resonators\cite{Dassonneville2020} and performing efficient state preparation\cite{BLOG_initialization} and embedded calibrations\cite{Vepsalainen2022}. Calibrations and optimization routines require interleaving quantum and classical processing and efficient execution of those could lead to dramatic improvements in attainable performance metrics\cite{Rol2017, Tornow2022}. Finally, benchmarking protocols require fast classical computation and randomness. Such hybrid quantum-classical processing is crucial to unleash the full potential of quantum advantage.

\setcounter{footnote}{0} 

Recently, several published manuscripts have addressed the requirements for quantum-classical processing (QCP) integration\cite{Corcoles2021, Lubinski2022}. Most of the effort, however, has been placed on gate-level definitions, which limit the scope of sequences and algorithms that can be expressed and executed on the quantum hardware. We     argue  that a deeper discussion of QCP at the pulse-level\cite{Smith2022} is important for several reasons\footnote{To control the dynamics of the quantum system at the physical level, the Hamiltonian is coupled to external control fields. These control fields are continuous functions of time, which is what we define as the \textit{waveform representation} of the quantum dynamics. The pulse-level representation is a representation that takes advantage of the fact that the waveforms could be chunked into \textit{pulses} which are shorter in time and allow for a more efficient description of the desired quantum dynamics.}. First, while gate-level is excellent for providing a unified birds-eye view for all types of Quantum Processing Units (QPUs), a great deal of performance can be extracted  from pulse-level abstractions at the current stage of quantum computing. Some examples include specifying the parameters of the employed QPU, optimizing pulse shapes and their timings, employing pulse-level dynamical decoupling methods, performing application-specific pulse-level optimizations such as unifying multiple gates into a small number of pulses (e.g., optimal control), performing application-specific pulse-level calibrations such as calibrating a specific multi-qubit gate, making use of not-discriminated measurement values, and more. Second, evaluation of hardware performance requires pulse-level control. Lastly, the gate-level abstraction does not allow separating the underlying system to its components and evaluating individual components behavior and performance. For example, it is impossible to create benchmarks that separate the quantum processor from the control hardware.

In this work, we categorize and demonstrate the requirements from QCP at the pulse-level and propose well defined benchmarks for the quantum control system, which is the key element for successful implementation of QCP. We utilize QUA\cite{web_QUA}, a native pulse-level language, as a framework for definitions and examples, making use of its expression of quantum operations concurrent with classical processing at all relevant timescales. It is important to note that QUA is used here only as a tool, and the suggested benchmarks are completely general.

\section{Quantum-classical processing requirements}  \label{CH2: reqs}


\begin{figure*}
  \centering
  \includegraphics[width=2.05\columnwidth]{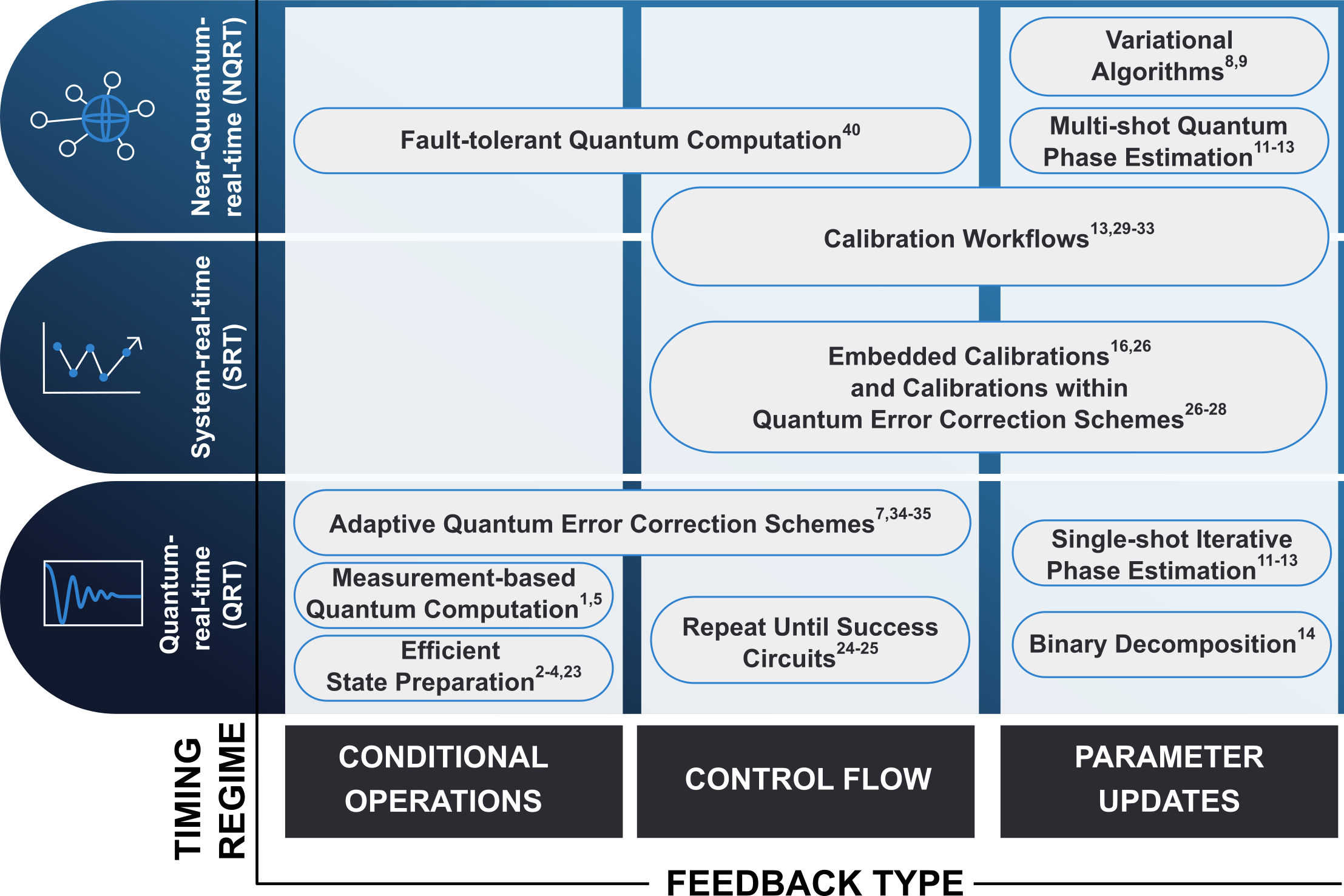}
  \caption{Examples of required functionalities and use cases that the ideal quantum controller should support, divided onto two conceptual dimensions: feedback types and timing regime for such feedback to be useful.}
  \label{fig:fig1}
\end{figure*}


We define \textit{quantum-classical processing} (QCP) as any computation in which \textbf{i)} a quantum device undergoes controlled dynamics; \textbf{ii)} measurements of the quantum device are performed; \textbf{iii)} classical processing is performed on the quantum measurement results; \textbf{iv)} the result of the classical processing modifies the subsequent dynamics of the quantum device. Note that in our definition, a full cycle of quantum-classical feedback must occur (i.e. including all four). 

We categorize the requirements from such QCP along two conceptual dimensions. The first one is the type of effect the classical processing has on the subsequent quantum processing. \\

\vspace{0.5cm}
We divide the feedback type into three categories:
\begin{itemize}[leftmargin=*]
\item \textcolor{QM_blue}{\textbf{\textit{Conditional operations}}}, where pulse-level operations are either enabled or disabled according to a Boolean result of classical processing of the measurement results, without affecting the timing of the program and without introducing "jump" (go-to) instructions to the flow of the program. 

\item \textcolor{QM_blue}{\textbf{\textit{Control flow}}}, where the decision regarding what branch, or subprogram, should be played is decided according to rules which rely on the classical processing of the measurement results. 

\item \textcolor{QM_blue}{\textbf{\textit{Parameters update}}}, where the parameters of a subsequent pulse-level operation are modified according to the results of the classical processing of the measurement results.   
\end{itemize}
 
The second conceptual dimension is the \textit{classical feedback  latency }requirements, i.e. how fast the classical part of the QCP loop needs to be, measured from the end of the quantum measurement until the dependent quantum control operation. For this latency to be well defined, the quantum-classical interface must be well defined, which we discuss in Section\,\ref{CH5: Benchmarks}. From here on, we refer to the classical feedback latency as simply the \textit{feedback latency}. We build upon the previous definitions\cite{BLOG_IBM_NRT} and discuss three categories, in each of which the feedback latency is compared to a different time scale:

\begin{itemize}[leftmargin=*]
\item \textcolor{QM_blue}{\textbf{\textit{Quantum-real-time (QRT)}}} describes the case in which the QCP feedback loop must be closed while at least part of the quantum system is undergoing coherent evolution. Therefore, QRT requires feedback latency significantly shorter (typically $\sim 1\%$ or less) than the lifetime of the QPU or some of its constituents. For the purpose of this work, we further require QRT to be time-deterministic, i.e., that it takes a deterministic amount of time. This means the timing of all control and measurement operations is completely determined by the program and its run-time inputs, including qubit measurement results. Deterministic timing of control and measurement operations is important, for example, to keep track of the evolution of the system during idle times, the dependency of certain control operations on phase\cite{Ofek2016}, etc. While there are cases of non-deterministic real-time operations, for simplicity we do not discuss them here.

\item \textcolor{QM_blue}{\textbf{\textit{System-real-time (SRT)}}}, describes the case in which the QCP feedback loop must be closed within a duration shorter than the dominant timescales of system parameter drifts (or at least part of it). To date, in all quantum computers, parameters of the system drift with time (e.g., qubit frequency, laser intensity, DC voltage biases across the system that tune various couplings, and other important parameters that affect the execution of a quantum program on the QPU). As long as the feedback latency is significantly shorter than the dominant timescales of these drifts, SRT processing can be used to re-calibrate and compensate for them, employing parameter updates, to improve overall system performance.

\item \textcolor{QM_blue}{\textbf{\textit{Near-quantum-real-time (NQRT)}}}, also referred to as near-real-time (NRT),   describes the case in which QCP feedback loop is desired to be as low as possible relative to the duration of the quantum circuit (including averaging). While the ability to run QRT and SRT processing fast enough is necessary in order to run the desired protocol (e.g., to compensate for a drift the SRT processing has to be shorter than the drift time or else the compensation will not work), the duration of NRT processing only affects run-time, and does not impact the ability to run the protocol nor its accuracy. NRT processing is important for many use cases, such as hybrid quantum-classical algorithms, calibration and optimization workflows, quantum error correction protocols, etc. In many of these cases, today, quantum computers are limited by classical communication and program loading times, which can be up to several of magnitude higher than the quantum processing (including the QRT and SRT). Note that IBM’s performance benchmark, CLOPS\cite{Wack2021}, is an example for an existing NRT benchmark. 
 \end{itemize}

Note that the actual feedback latencies required in the above categories vary significantly between various qubit and QPU types as the time scales they are compared to vary significantly. Coherence times, relevant for QRT, can vary between milliseconds in superconducting qubits, to several seconds in trapped ions qubits. Significant parameter drifts timescales, relevant for SRT QCP, vary between hundreds of nanoseconds , in the case of crystal defects, to many hours in the case of thermal variations.  Finally, the quantum operation times relevant for NRT also varies dramatically from nanoseconds in superconducting and spin qubits, $100\,s$ of nanoseconds for neutral atom based qubits, all the way to $100\,s$ of microseconds for trapped ions qubits.  

\begin{figure*} 
  \centering
  \includegraphics[width=2\columnwidth]{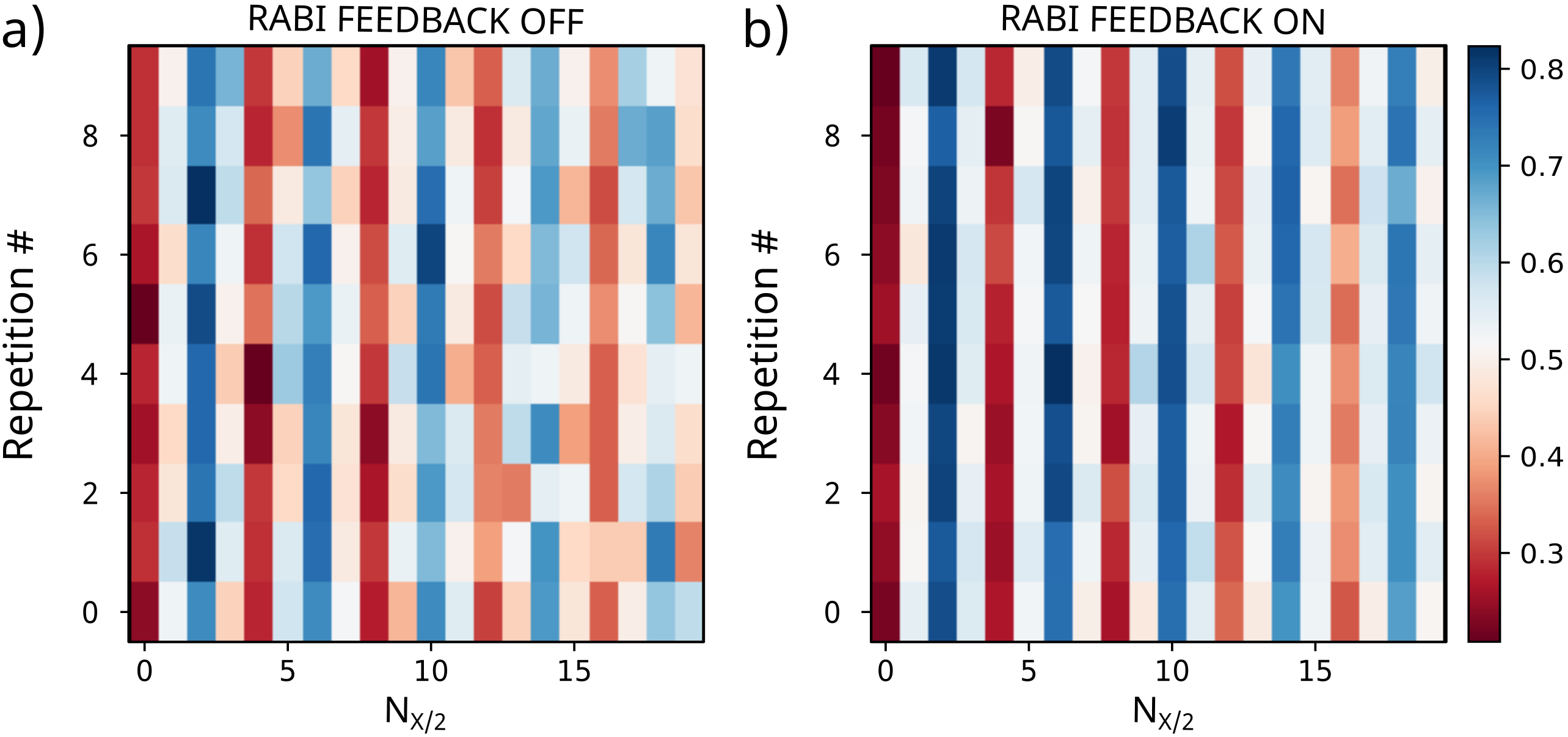}
  \caption{Rabi oscillations discretized into $\pi/2$ rotations and repeated 10 times to observe instability, with feedback-based re-calibration of the Rabi rate either not applied \textbf{a)} or applied \textbf{b)}. The Rabi rate corrections are performed in system-real-time using the OPX+\cite{web_OPX}. Figure courtesy of Gilbert et al.}
  \label{fig:fig2}
\end{figure*}

Figure \ref{fig:fig1} shows the above categorization and examples of use cases in each category. Protocols such as efficient state preparation via conditional pulses\cite{Hoyer2005, Devulapalli2022,Bravyi2006, foss2023} or repeat-until-success\cite{Paetznick2014, Moreira2022}, measurement-based quantum computation\cite{Piroli2021, Raussendorf2001}, and single-shot iterative phase estimation\cite{Corcoles2021, Granade2022, Lubinski2022} all require mid-circuit measurement and classical feedback latency that is done within a single \textit{shot}, i.e., a sequence of control and measurement after which all qubits are reset. Embedded calibrations\cite{Vepsalainen2022, Kelly2016, Wagner2021, Wagner2022} and some calibration workflows\cite{Lubinski2022, Xu2022, McEwen2021, Werninghaus2021, Klimov2020, Proctor2020}, which allow retuning of system parameters before significant drifts require SRT QCP. Finally, variational algorithms\cite{Cerezo2021, Bharti2022}, and non-Clifford gates via magic state distillation for fault tolerant quantum computation require NRT, while adaptive quantum error correction schemes are QRT\cite{Ryan2021, Brien2017, Tansuwannont2022}.  
Note that we make a distinction between single-shot and multi-shot iterative phase estimation. While in multi-shot iterative phase estimation after each measurement the eigenstate of interest is re-prepared to measure a different bit in the binary representation of the phase, in the single-shot protocol QRT feedback is used to skip the re-preparation of the phase sate making the phase estimation more efficient\cite{Corcoles2021, Lubinski2022}. 
In Section\,\ref{CH3: examples}, we present recent implementation of some of these use cases and define control benchmarks that allow assessing the performance in such use cases.

\section{Quantum-Classical Processing Use Cases Examples }  \label{CH3: examples}

We now review and discuss a series of published and unpublished examples that emphasize some of the essential requirements of QRT, SRT and NRT quantum classical processing.

\subsection{SRT parameters update and embedded calibrations: tracking of qubit frequency and Rabi rates}

In the work of Gilbert et al.\cite{Gilbert2023} the authors demonstrate the all-electrical control of spin qubits in silicon, which allows for increased qubit density and reduced sensitivity to electrical noise. This is enabled by the ability to bias quantum dots to a point where two orbital states become degenerate, thus making spin-orbit interactions significant. In turn, high electric dipole spin resonance (EDSR) Rabi rates are achieved. However, the Rabi rate are then highly dependent on the gate voltages and therefore, fluctuations in the electrostatic environment can lead to deterioration of single qubit gate fidelities. Evidently, by measuring the probability of a qubit flip after an odd number of applied $\pi /2$ pulses the authors demonstrated discrepancies from the ideal $P_{flip}$ of $0.5$. The sensitivity of the Rabi rate to the environment can then be mitigated by using SRT QCP to perform measurement-based re-calibration of the voltage. Figure\,\ref{fig:fig2} shows an example of such re-calibration (similar to Extended Data Fig.5 of Ref.\cite{Gilbert2023}) performed by correcting the microwave amplitude of a factor: \vspace{0.1cm}
\begin{equation}
    \centering
    \delta A_{MW} = -gain \cdot (P_5 - P_7 + P_9 - P_{11}),
\end{equation}
 
where $P_n$ is the measured spin flip proportion after $n$ rotations. The use of such a protocol was shown by the authors to be instrumental in obtaining a single-qubit gate fidelity of $99.93\,\%$.

\begin{figure*}
  \centering
  \includegraphics[width=2\columnwidth]{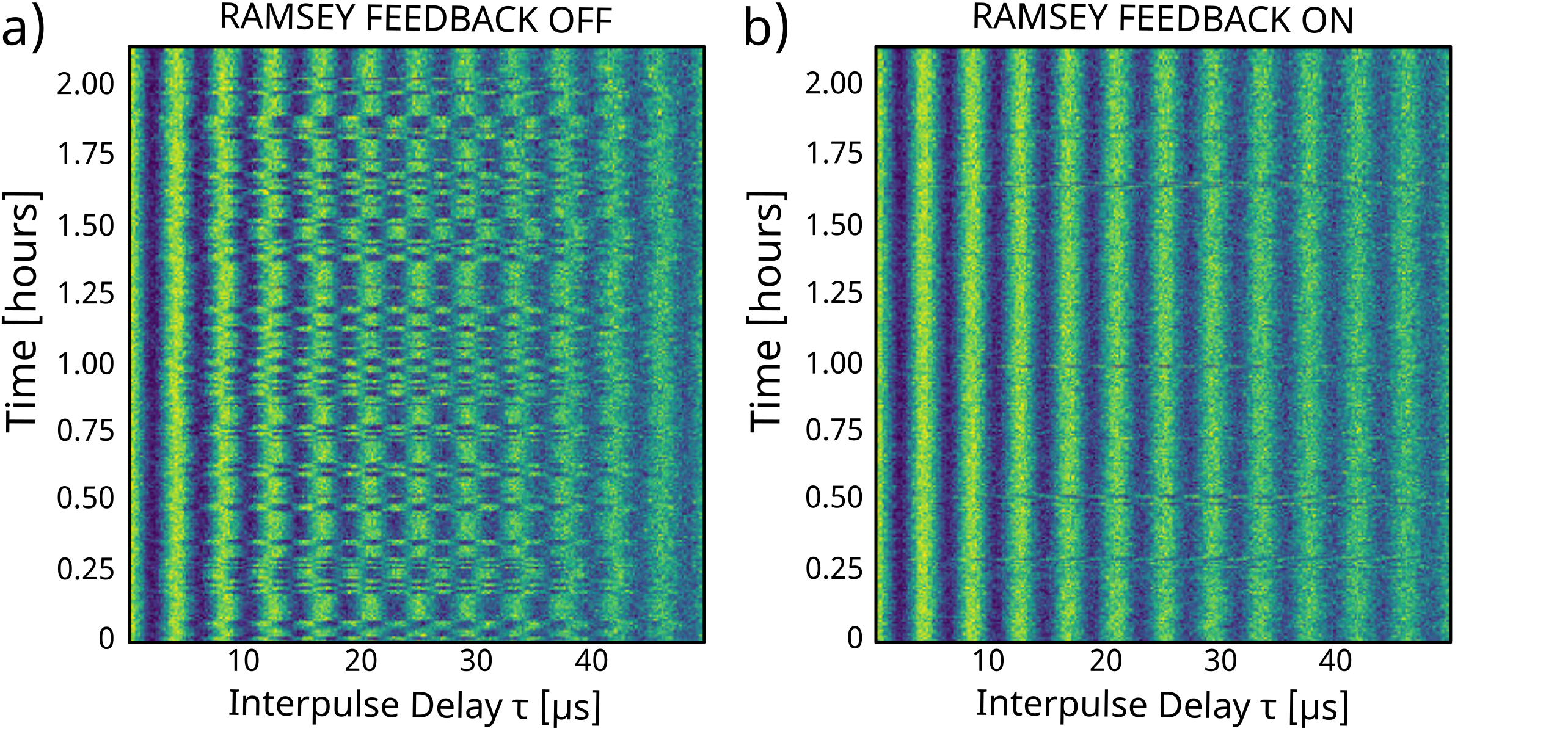}
  \caption{Ramsey scans performed thanks to an OPX+\cite{web_OPX} with system-real-time frequency tracking either \textbf{a)} applied or \textbf{b)} not applied. The correction is based on a few-points Ramsey measurement embedded within the circuit. Figure courtesy of Prof. David Schuster. }
  \label{fig:fig3}
\end{figure*}

The amplitude of pulses is only one example of a parameter that can be modified by SRT QCP. Figure\,\ref{fig:fig3}, courtesy of Prof. David Schuster, shows a SRT re-calibration of the resonant frequency of a transmon qubit\cite{Vepsalainen2022, Agrawal2022}. In this sequence, repeated Ramsey measurements are performed at different frequencies to retrieve the frequency drift. The quicker the re-calibration protocol is, the wider the bandwidth of frequency fluctuations that can be suppressed. Here, the feedback latency was on the order of a microsecond, which was significantly lower than the Ramsey sequences themselves, hence not limiting the re-calibration bandwidth. To verify the effectiveness of the compensation, a Ramsey sequence is run repeatedly for several hours, without (with) interleaved compensation, showing the resulting instability (stability), as shown in Figure\,\ref{fig:fig3}a (b). 

Since in order to comply with the short timescale constraints, such re-calibrations must typically run entirely within the dedicated control hardware that runs the desired application, we call them \textit{embedded calibrations}. 
Clearly, the speed at which the required measurements, SRT processing and feedback (i.e., the tracking and updating the relevant parameter) is performed determines the performance of the re-calibration. 
The re-calibration bandwidth, in the best case, is limited to $1/(8\cdot\tau_{tot})$
$[Hz]$, where $\tau_{tot}$ 
is the total latency for the required correction, including measurements, classical calculation and feedback, etc. 
In this context we want the control hardware latency to be negligible, for the correction bandwidth to be limited by factors related to the observed system.

\subsection{QRT parameter update and control flow: quantum error correction and Fock-state binary decomposition}

In addition to parameter updates, the timing and latency of classical operations should also be considered in the context of general control flow, e.g., when complicated branching and logic statements are evaluated during the sequence, to impact the sequence dynamically. 

One of the seminal works involving real-time control is the work by Ofek et al.\cite{Ofek2016}. Here, a logical state encoded in a cat code qubit, implemented with a 3D bosonic cavity, was maintained by applying repeated photon parity measurements with an ancilla transmon qubit. This allowed for a quantum error correction protocol, detecting parity-altering photon number changes, which correspond to phase errors in the logical space. The experimental procedure consisted of encoding the logical state in the cavity mode, performing repeated rounds of parity measurements, and finally decoding the state based on the processing of the parity measurements. 

Such a state preservation algorithm required both QRT branching as well as QRT parameter updates. Branching was required to maintain the ancilla in the ground state for as long as possible, which improves its measurement and gate fidelity. Real-time parametric phase updates were required to correct Kerr-induced phase jumps, which depend on the time of their occurrence. Lastly, the decoding stage required real-time ancilla phase shifts to be applied based on the number of parity changes observed during the preservation time.  

\begin{figure*}
    \includegraphics[width=2.05\columnwidth]{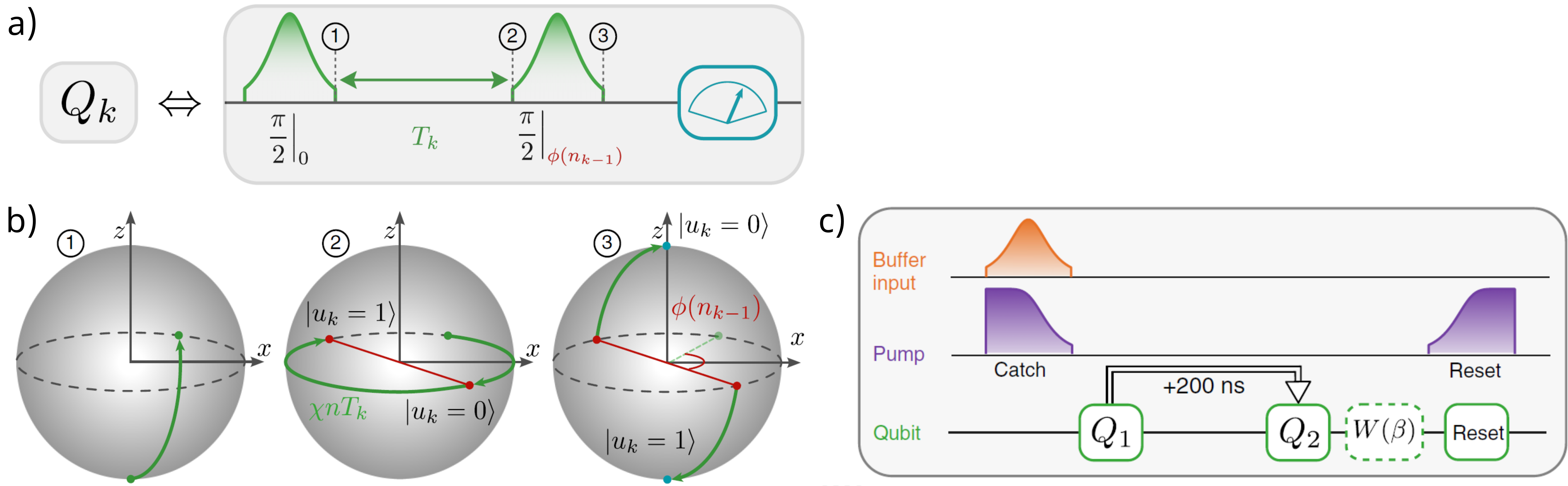}
    \caption{\textbf{a)} Pulse sequence to perform binary decomposition with \textbf{b)} relative Block sphere representations of the qubit trajectories. \textbf{c)} Pulse sequence and feedback representation to perform the binary decomposition protocol, where the qubit performs photo counting bit by bit with $Q_K$ pulse sequences where $Q_2$ utilizes the information obtained in $Q_1$. The feedback latency of such an operation is kept around $200\,ns$. Figure adapted from Dassonneville et al.\cite{Dassonneville2020} with permission from the publisher.\label{fig:fig4}}
\end{figure*}

Another example of real-time control flow was demonstrated in the work of Dassonneville et al.\cite{Dassonneville2020}. The team demonstrated a single-shot photon counter that can resolve the number of photons in a microwave pulse up to three microwave photons (i.e., $N \, mod \, 4$), by using a transmon qubit. The counting is done bit by bit, in a sequence of conditional gates which maps the cavity photon number parity to the qubit’s state. The protocol, shown here in Figure\,\ref{fig:fig4}, adapted from Dassonneville et al.\cite{Dassonneville2020}, requires rotating the qubit drive frame by an angle that depends on the result of all previous digits in real-time. When generalized to resolving a greater number of photons, i.e., the binary decomposition of the number of photons requires many digits, the frame rotation resolution increases, making it inefficient to do with control flow and requiring parameters updates\cite{Lubinski2022}. Finally, beyond the number-state measurement, Dassonneville et al. also employed QCP for a repeat-until-success-based qubit initialization, demonstrating control-flow and showing how different categories of the above requirements coexist in the same application.   

The number-state measurement protocol resembles the iterative phase estimation algorithms mentioned in Figure\,\ref{fig:fig1}. Two additional demonstrations of these important algorithms were recently done by Corocles et al.\cite{Corcoles2021} and Lubinski et al.\cite{Lubinski2022}. In both cases, a two-qubit system was used. The target qubit was prepared in an eigenstate of a controlled \textit{U} operator, and the auxiliary qubit was used to infer the phase of the eigenstate via repetitive applications of the controlled \textit{U} gate. In the case of Ref.\,\cite{Corcoles2021}, only the auxiliary qubit was reset between phase measurement rounds, which makes it a single-shot phase estimation procedure (see Figure\,\ref{fig:fig1}). In the case of Ref.\,\cite{Lubinski2022}, the target qubit was reset once every 8 shots, making it an intermediate case between single shot and multi-shot phase estimation. Both implementations, however, required real time parameter updates to be applied during the coherence time of another qubit, making them QRT type feedback experiments.

\subsection{QRT conditional pulses: Multi-qubit active state preparation}

While in some cases, full control flow (i.e., branching) is necessary, in other cases conditional operations are sufficient. One clear example is active state preparation, where to reset a qubit to the $|0>$ state, it is first measured and then a $\pi$-pulse is played conditioned on the measurement result. To be useful, this must occur on a timescale significantly shorter than the qubit relaxation time.  

The work of Zhang et al.\cite{Zhang2023} demonstrates a simultaneous active reset of 10 qubits. The authors perform multiplexed single-shot readout and apply a $\pi$-pulse on the qubits measured in the $|1>$ state. Such real-time operations enable both a high-repetition rate of the experiment\,(as it eliminates the need to wait for the qubits to naturally decay) and high initialization fidelity, which is critical due to the demanding statistical requirements of many qubit state analysis. In fact, the active reset reduced the initialization time from milliseconds in the case of thermalization-based state preparation, to microseconds, while simultaneously achieving superior fidelities. The preparation fidelities obtained by thermalization on a similar setup are estimated to be $\sim89\,\%$\cite{Mirhosseini2019}. Instead, the authors showcase a reset fidelity on the order of $94\,\%$ and propose how to increase it further.  
Unlike methods such as restless tune-up\cite{Rol2017}, this protocol does not place any constraints on the circuit and is less sensitive to the quantum-non-demolition\,(QND) nature of the measurement. Note that other initialization techniques exist and come with varying requirements on the controller’s QRT or NRT capabilities\cite{Dassonneville2020, web_QUA}.

Each of the above examples, and many others proposed in recent years, motivate the need for extremely low feedback latencies of QCP, which often necessitate to be performed “close to the QPU”. In fact, in all of the above examples, the QCP protocols were performed entirely from the \textit{quantum controller} - the device that dynamically orchestrates pulse-level quantum control operations. An advanced quantum controller that can execute classical processing operations is therefore key to allow for the stringent QCP latency requirements\,\cite{web_OPX}. It is for this reason that in the following sections we focus on the quantum controller and its benchmarks.

\section{Programming at the pulse-level }  \label{CH4: QUA}

To define quantum control benchmarks, we utilize QUA\cite{web_QUA}, a comprehensive pulse-level language for quantum-classical programming, which was designed to encompass the entire range of requirements described here. It is important to note that QUA is used here only as a tool to rigorously define the required behavior of the controller in each benchmark. In fact, in places where it serves the purpose of the benchmarks definitions and their clarity we allow ourselves to simplify the syntax even if it is not fully consistent with current QUA specifications. Therefore, the benchmarks are completely independent of the pulse-level language and can serve as a general tool to assess controller’s performance in QCP.  

At its heart QUA integrates the following two concepts: 
\begin{itemize}[leftmargin=*]
    \item Pulse-level quantum control and measurement instructions that can be programmed with precise timing.  
    
    \item General classical processing that can be fed by quantum measurements or other inputs and affect pulse-level instructions parameters and control flow. 
\end{itemize}
We briefly introduce QUA with an example. 

\begin{lstlisting}[language=Python, label={code1:example}, caption= QUA example code of a Ramsey experiment. \textbf{\textcolor{QM_green}{Green}}: elements of the configuration file. \textbf{\textcolor{QM_blue}{Blue}}: real-time classical variables. \textbf{\textcolor{QM_red}{Red}}: pulse-level commands. \textbf{\textcolor{QM_orange}{Orange}}: constants. \textbf{\textcolor{QM_purple}{Purple}}: control flow statements.] 

tau_max = 400
n_avg = 200
max_latency = 300

int n
int tau 
fixed x
fixed discr_threshold = 0
bool state
bool[tau_max] states
int freq_correction
int qubit_freq
int i

for (n=1, n<n_avg, n+1):  
   i = 0
   for (tau, 0, tau<tau_max, tau+1): 
      strict_timing: 
       
         #Active Reset 
         measure(readout_pulse, ...
                 resonator, demod(x))   
         state = x > discr_threshold 
         wait(max_time= max_latency,qubit)  
         if(state): 
             play(pi, qubit)
                
         #Ramsey Sequence 
         play(pi_half, qubit)				          
         wait(tau, qubit)
         play(pi_half, qubit) 
         align(qubit, resonator) 
         measure(readout_pulse, ...
                 resonator, demod(x))    
         state = x > discr_threshold 
         states_vec[i++] = state 
         
   freq_correction = ...
            correction_fn(states_vec) 
   qubit_freq = qubit_freq+freq_correction 
   update_frequency(qubit, qubit_freq) 
\end{lstlisting}

The QUA code in Listing\,\ref{code1:example} describes a Ramsey experiment performed on a superconducting qubit device containing a single qubit and a readout resonator. There is a nested loop, where the inner loop scans the Ramsey sequence delay, and the outer loop is used for repeating the delay scan many times to collect statistics. The code includes an active qubit reset at the beginning of every Ramsey shot\,(that should be done in QRT) as well as compensation for the drift of the qubit frequency after every scan of the Ramsey sequence delays\,(that should be done in SRT).  To simplify the code, the actual expressions of the classical processing that are needed to correct for the frequency of the qubit based on the previous measurements are wrapped in a macro $\textbf{correction\_fn()}$. Here, this macro is given just for illustration. A real frequency correction example, and other similar SQR protocols, are discussed in Section\,\ref{CH3: examples}. 

Note that QUA has general classical variables (blue in the code), which can affect both the control flow (e.g., the $\textbf{\textcolor{QM_purple}{if}(\textcolor{QM_blue}{state})}$ statement) as well as the parameters of the pulse-level operations (e.g. the $\textbf{\textcolor{QM_red}{wait}(\textcolor{QM_blue}{tau}, \textcolor{QM_green}{qubit})}$ and the $\textbf{update\_frequency(\textcolor{QM_green}{qubit}, \textcolor{QM_blue}{qubit\_freq})}$ statement). For the basic usage of the sequence \textbf{\textcolor{QM_purple}{if}(\textcolor{QM_blue}{state})$\rightarrow$\textcolor{QM_red}{play}()}, one may also use the conditional play statement \textbf{\textcolor{QM_red}{play}(\textcolor{QM_green}{pulse},\textcolor{QM_green}{qubit}, condition = \textcolor{QM_orange}{bool\_value})}. In addition, the code that is wrapped in a \textbf{strict\_timing} statement enforces all the pulse-level operations (in bold red, e.g., $\textbf{\textcolor{QM_red}{play}}$, $\textbf{\textcolor{QM_red}{wait}}$, $\textbf{\textcolor{QM_red}{measure}}$) in it to be performed “back-to-back”, with no “time gaps”, even when there is a dependency on classical processing or a dependency on previous measurements. In case of a feedback latency that prevents back-to-back operations, the user can add the latency that is required to “fill the gap” using a $\textbf{\textcolor{QM_red}{wait}}$ statement. To make the code simpler, and in cases in which time-determinism is not important, we also allow using a $\textbf{\textcolor{QM_red}{wait}}$ statement with a $\textbf{\textcolor{QM_blue}{max\_time}}$ parameter, which instructs the controller to wait the minimal amount of time needed for the feedback, up to the given time limit. Thus, $\textbf{\textbf{strict\_timing}}$ allows to enforce timing and therefore expresses time-deterministic quantum-classical processing. Outside of the $\textbf{\textbf{strict\_timing}}$ block, the code does not enforce timing, but rather requires that operations happen as soon as possible, given their dependency on previous operations in the code. In addition to the local processing, QUA variables may be saved and streamed out to remote classical computation resources. The ability to stream the data during the QUA program execution allows pipelining of QRT, SRT and potentially NRT calculations performed on the control with NRT calculations that require additional compute resources. It is the combination of (1) classical computation, (2) pulse-level quantum operations, and (3) the ability to enforce strict timing , that allows QUA to cover the entire range of QCP use-cases and makes it suitable for defining quantum control benchmarks at the pulse-level. 

\section{Pulse-level quantum-classical benchmarks for quantum controllers }  \label{CH5: Benchmarks}

\setcounter{figure}{4}
\begin{figure*}
  \centering
  \includegraphics[width=2.04\columnwidth]{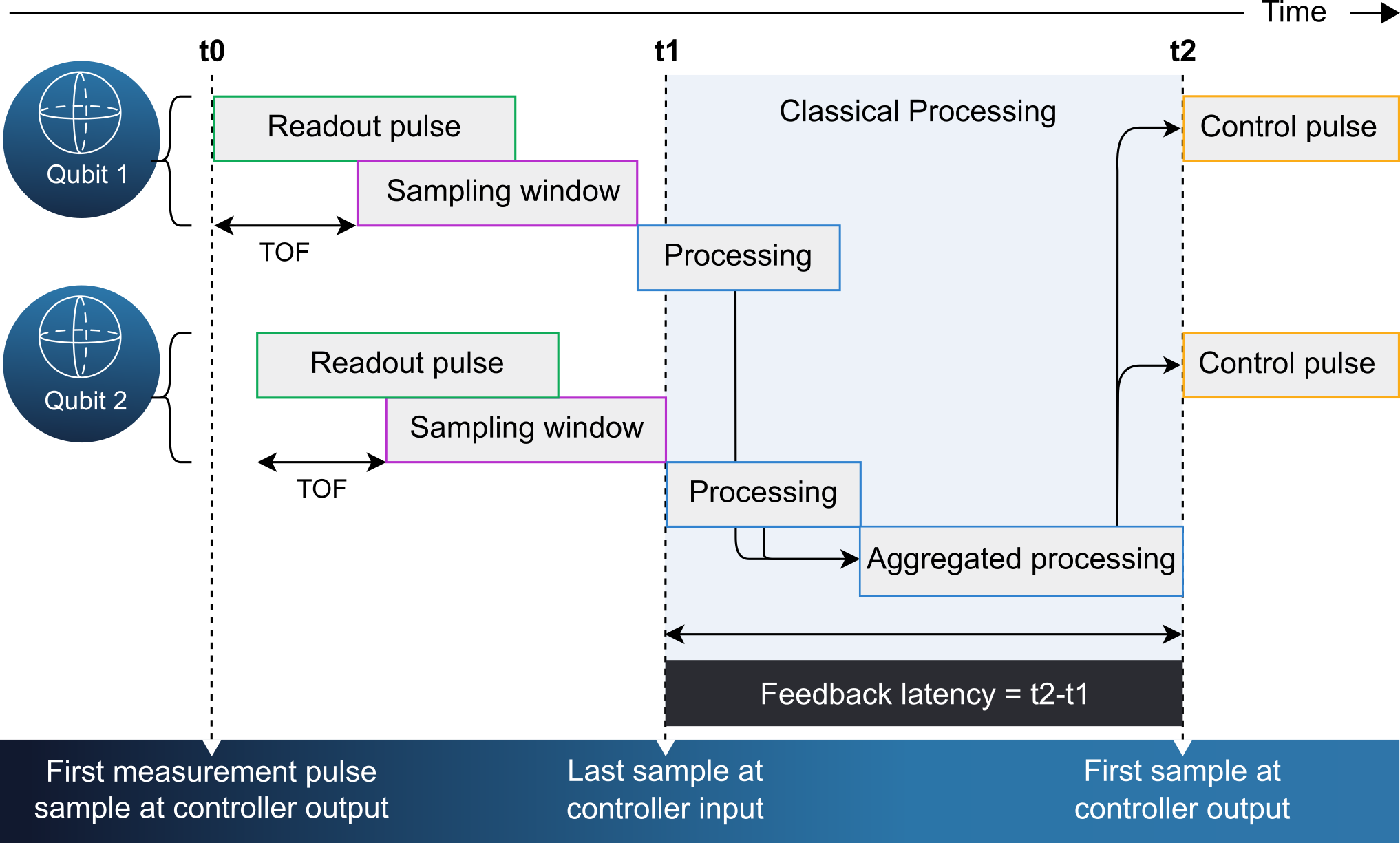}
  \caption{Timeline of controller operations for a generalized feedback sequence. The controller feedback latency is defined as the time from the last sample required for the calculation that affects the output is sampled until the first sample of the dependent pulses is sent at the controller output. An output pulse may depend on multiple inputs. In such a case, the latency is defined as the time of the last sample of all inputs that impacts the outputs. One method to measure the time of the last sample required for the calculation is to sample the time of measurement pulse and add the time-of-flight (TOF) and sampling window length. }
  \label{fig:fig5}
\end{figure*}

As discussed above, the key component that must be designed and implemented to meet the QCP requirements is the quantum controller. In this section, we define and describe a comprehensive suite of benchmarks for the quantum controller. We focus on two categories of benchmarks: benchmarks for QRT use cases where classical feedback is done during a single shot and benchmarks for SRT use cases where classical feedback is done after multiple shots based on their measurements. While the second category can be relevant for some NRT use cases as well, the classical processing we enforce is relatively light and does not represent the requirements for some important NRT use cases (e.g., error decoding). We leave benchmarks that enforce heavy classical processing for future work. The proposed benchmarks isolate the controller from the rest of the system and allow componentization of the stack that should help compare controllers (or controllers’ components) and build best-of-breed solutions. 

\subsection{Benchmarks definitions}

All the controller benchmarks defined below essentially measure feedback latencies in a well-defined test. We define this latency to be the duration from the time the controller samples the last qubit measurement signal on which the classical processing depends, until the first sample of the response signal that is sent from the controller to the qubits (see Figure\,\ref{fig:fig5}). For this to be well-defined, the interface between the controller and the QPU must be well-defined. Here we define it as the analog interface of the digital-to-analog converters and analog-to-digital converters of the control system. The analog components chain that is between these interfaces and the physical qubits varies between different QPUs, however, the propagation delay through it is typically relatively negligible. In cases of optical based measurements that are done with either photodetectors or cameras, we measure from the optical-to-digital interface.    

After the controller samples the input signal, it must process it to infer qubit estimated state. There are various processes that can be used depending on the qubit type, such as demodulation and weighted integration (e.g., SC and spin qubits), time-tagging or counting of TTL pulses (e.g., trapped ions and NV centers), and image processing (e.g., neutral atoms and trapped ions). For the benchmarks to be well defined, the exact process must be specified. Here we use a demodulation and weighted integration process\cite{Corcoles2021} defined in Appendix\,\ref{APP3} . It is also important to enforce the requirements on the pulse that the controller applies in response to the classical feedback. Here we enforce a pulse with a Gaussian envelope that is $20\,ns$ long and an intermediate frequency of $100\,MHz$, defined more precisely in Appendix\,\ref{APP3}.

Finally, each of the benchmarks defined below can also depend on various parameters. For example, there can be more than one quantum measurement generating inputs for the classical processing and more than one quantum operation that is affected by the processing result. We denote this as “fan-in/fan-out” and discuss this further in the Appendix\,\ref{APP1}. Additionally, multiple experimental parameters could be updated as a result of the classical processing in case of parametric updates. In this section, we show each of the benchmarks in its simplest case and propose additional variations in Appendix\,\ref{APP1}.

We define the following categories of benchmarks: \textcolor{QM_blue}{\textbf{QRT benchmarks}} and \textcolor{QM_blue}{\textbf{SRT Benchmark}}.

The first category is designed to measure the feedback latency of the building blocks in important QRT use cases (conditional gates for teleportation, state preparation using repeat-until-success, efficient iterative phase estimation, etc.). The second category is designed to measure the latency in cases where relatively inexpensive classical processing is done on the results of many shots, to update control parameters of subsequent control and measurement shots. This benchmark is important to assess the ability of the controller to perform calibrations to overcome and correct drifts and thus measured SRT performance. In some cases, this benchmark can also assess the ability of the controller to perform a hybrid algorithm.

\subsection{QRT (Single-Shot) Benchmarks} \label{CH5.2: QRT Benchmarks}

The benchmarks in this section measure the controller performance when the timing is critical and deterministic. The benchmarks measure the feedback time between a measurement and the subsequent pulses they affect within a single experiment shot and hence are called Single-Shot benchmarks. We define three benchmarks designed to measure the controller’s ability to perform well in use cases requiring QRT QCP:
\begin{itemize}[leftmargin=*]
\item \textbf{BM1.1: conditional operations}
\item \textbf{BM1.2: control flow}
\item \textbf{BM1.2: parametric updates}
\end{itemize}

In the following, we define the simplest variations of these benchmarks, e.g., where the feedback is between a single input and a single output. In Appendix\,\ref{APP1}, we present a more comprehensive list of feedback processes that account for the general case of multiple inputs and outputs, as well as distributed vs. aggregated processing, etc.

\subsection*{BM1.1: Deterministic conditional operations } \label{BM1.1}

This benchmark measures the feedback latency for pulses that are played conditionally on processed measurement results (see Figure\,\ref{fig:fig5}). In this scenario, each measurement channel generates a number which is then compared to a threshold to produce a single bit, and a subsequent operation is performed conditionally on this bit. This test, in its simplest version, is defined in QUA as shown in Listing\,\ref{code2:BM1.1}. The full parametrization of this test is discussed in Appendix\,\ref{APP1}. The deterministic conditional operations benchmark is relevant to use cases such as the simultaneous 10 qubits active reset, demonstrated by Zhang et al.\cite{Zhang2023} (see Section\,\ref{CH3: examples}), teleportation\cite{Devulapalli2022}, etc.

We use the \textbf{timestamp}\footnote{The \textbf{timestamp} feature gives you, after execution, the exact time at which a command was executed.} feature  to measure the time in which quantum operations occur. Note that the calculation of the feedback latency in the code includes the subtraction of the $\textbf{\textcolor{QM_blue}{time\_of\_flight}}$ and the $\textbf{\textcolor{QM_blue}{sampling\_window}}$. This is because  $\textbf{\textcolor{QM_blue}{re\_time}}$ represents the timestamp at which the first sample of the $\textbf{\textcolor{QM_green}{readout\_pulse}}$ comes out of the controller output, while the time needed for the feedback latency measurement is the timestamp at which the controller samples the last sample of the returning pulse which affects the measurement result at the input. This timestamp can be calculated by taking $\textbf{\textcolor{QM_blue}{re\_time}}$ and adding to it the time it takes the readout pulse to arrive to the input of the controller ($\textbf{\textcolor{QM_blue}{time\_of\_flight}}$) and the $\textbf{\textcolor{QM_blue}{sampling\_window}}$, as depicted in  Figure\,\ref{fig:fig5}.

\begin{figure*}
\begin{lstlisting}[language=Python, label={code2:BM1.1}, caption= QUA code defining the BM1.1 benchmark for QRT conditional operations. Color coding same as in Listing\,\ref{code1:example}.] 
def benchmark_deterministic_QRT_conditinal_operation():
   fixed x 
   bool s

   strict_timing:
      measure(readout_pulse, readout_element, demod(x), timestamp-> re_time) 
      s = x > 0
      wait(max_time=max_latency, control_element)
      play(control_pulse, control_element, condition = s, timestamp-> ce_time)

   return feedback_latency = ce_time - (re_time + sampling_window + time_of_flight)
\end{lstlisting}
\end{figure*}

\subsection*{BM1.2: Deterministic control flow} \label{BM1.2}

\begin{figure*}
\begin{lstlisting}[language=Python, label={code3:BM1.2}, caption= QUA code defining the BM1.2 benchmark for QRT control flow. Color coding same as in Listing\,\ref{code1:example}.] 
def benchmark_deterministic_QRT_control_flow():
   fixed x 
   bool s
   s = False

   strict_timing:
      while s == False:
         measure(readout_pulse, readout_element, demod(x), timestamp-> re_time)
         s = x < 0   	
         wait(max_time=max_latency, readout_element)  				
      align(control_element, readout_element)
      play(control_pulse, control_element, timestamp->ce_time)      
   		
   return feedback_latency = ce_time - (re_time + sampling_window + time_of_flight)
\end{lstlisting}
\end{figure*}

This benchmark measures the feedback latency for the case where a jump in the flow of the program occurs based on the result of classical processing performed on one or more measurement results. In this test, we use the most relevant scenario for quantum computing, which is \textit{repeat-until-success}\cite{Paetznick2014, Moreira2022}.  The benchmark in its simplest version is defined in Listing\,\ref{code3:BM1.2}, with more variations detailed in Appendix\,\ref{APP1}. The deterministic control flow benchmark is relevant to use cases such as active qubit reset based on repeat-until-success, demonstrated by Dassonneville et al.\cite{Dassonneville2020} (see Section\,\ref{CH3: examples}), similar protocols needed for magic state distillation\cite{Litinski2019} and flag qubits based QEC\cite{Ryan2021}.

\subsection*{BM1.3: Deterministic parametric updates}  \label{BM1.3}

This benchmark measures the feedback latency for updating a parameter of one or more output pulses based on the result of classical processing performed on one or more measurement results. Here we take full advantage of the fact that we are working directly in pulse-level by specifying the concrete pulse parameters we want to modify, as opposed to more abstract quantities such as gate rotation angles. In Listing\,\ref{code4:BM1.3}, we define two benchmarks from this family, both rotate the phase of a qubit control pulse by an angle that depends on previous measurements and classical processing. In the first benchmark, a single measurement and state discrimination is performed, and the rotation angle is chosen from a look up table (LUT) of two values corresponding to the two possible states of the qubit. This benchmark is relevant to use cases in which certain pulse parameters can be optimized based on the result of mid-circuit measurements, for instance to compensate for state dependent phase shifts\cite{Ofek2016}. In the second benchmark, a series of 16 measurements is performed and the rotation angle is determined by the binary representation of the 16 state estimations. This benchmark is relevant to use cases such as the single-shot iterative phase estimation\cite{Corcoles2021, Granade2022, Lubinski2022} and the binary decomposition\cite{Dassonneville2020}. A variety of other deterministic parametric updates are defined in Appendix\,\ref{APP1}. 

\begin{figure*}
\begin{lstlisting}[language=Python, label={code4:BM1.3}, caption= QUA code defining the BM1.3 benchmark for QRT parametric updates. Color coding same as in Listing\,\ref{code1:example}.] 
def benchmark_deterministic_QRT_parametric_update_frame_LUT ():
    fixed x
    int s
    fixed frame_rot_ang
    fixed[2] frame_lut = [0.1, 0.2]
    strict_timing:
        measure(readout_pulse, readout_element, demod(x), timestamp-> re_time)
        s = x > 0    
        frame_rot_ang = frame_lut[(s)]     #returns 0.1(0.2) when S=0(1)
        wait(max_time=max_latency, control_element)
        frame_rot_2pi(frame_rot_ang, control_element)  
        play(control_pulse, control_element, timestamp -> ce_time)

    return feedback_latency =ce_time - (re_time + sampling_window + time_of_flight)


def benchmark_deterministic_parametric_update_frame_binary_rep():
   fixed[16] x
   bool[16] s
   fixed frame_rot_ang 

   strict_timing:
      for (i, 0, i < 16, i+1):
         measure(readout_pulse, readout_element, demod(x[i]), timestamp-> re_time[i])
         s[i] = x[i] > 0
      frame_rot_ang = bin2dec(s)/2**16       
      wait(max_time=max_latency, control_element)
      frame_rot_2pi(frame_rot_ang, control_element)
      play(control_pulse, control_element, timestamp-> ce_time)

   return feedback_latency = ce_time - max(re_time + sampling_window + time_of_flight)
\end{lstlisting}
\end{figure*}

\subsection{SRT (Multi-shot) Benchmark}

\subsection*{BM2.1: Multi-shot parameter updates} \label{BM2.1}

\begin{figure*}
\begin{lstlisting}[language=Python, label={code5:BM2.1}, caption= QUA code defining the BM2.1 benchmark for SRT multi-shot parametric updates. Color coding same as in Listing\,\ref{code1:example}.] 
def benchmark_deterministic_SRT_parametric_update(N_in=10,N_out=10,N_shots=1000): 
   bool[N_in] s 
   fixed[N_out][2**N_in] 
   T = random(N_out, 2**N_in) 
   fixed[2**N_in] H      	 
   fixed[N_out] f 
   int[N_in] x 
   for (i, 0, i < 2**N_in, i+1):  
      H[i] = 0 
   strict_timing: 
      for (i, 0, i < N_shots, i+1): 
         for i in range(N_in):  # meta-programming - loop unrolled
                                # measurements are parallel
            measure(readout_pulse, readout_element[i], demod(x[i]), ...
                    timestamp->re_time[i]) 
            s[i] = x[i] > 0          
         H[bin2dec(s)]++ 
   f += T*(H-H0) 
   for i in range(N_out):       # meta-programming - loop unrolled
      update_frequency(control_element[i], f[i])   
   strict_timing: 
      for i in range(N_out):    # meta-programming - loop unrolled
         play(control_pulse, control_element[i], timestamp-> ce_time[i]) 
   return feedback_latency= ce_time[0]-max(re_time + time_of_flight + sampling_window)  
   # All the control elements are aligned, thus we use the timestamp of element 0
\end{lstlisting}
\end{figure*}

Here we define a benchmark, shown in Listing\,\ref{code5:BM2.1} for cases in which relatively inexpensive classical processing is done on the results of many shots, to update control parameters of subsequent control parameters. This is important in order to assess the ability of the controller to perform calibrations to overcome and correct drifts and thus measured SRT performance. In some cases, this benchmark can also assess the ability of the controller to perform variational algorithms. Prototypical use cases for this type of program flow and processing are found in Refs.\cite{Kelly2016, Wagner2021, Proctor2020}. 

\begin{figure*}
  \centering
  \includegraphics[width=2.04\columnwidth]{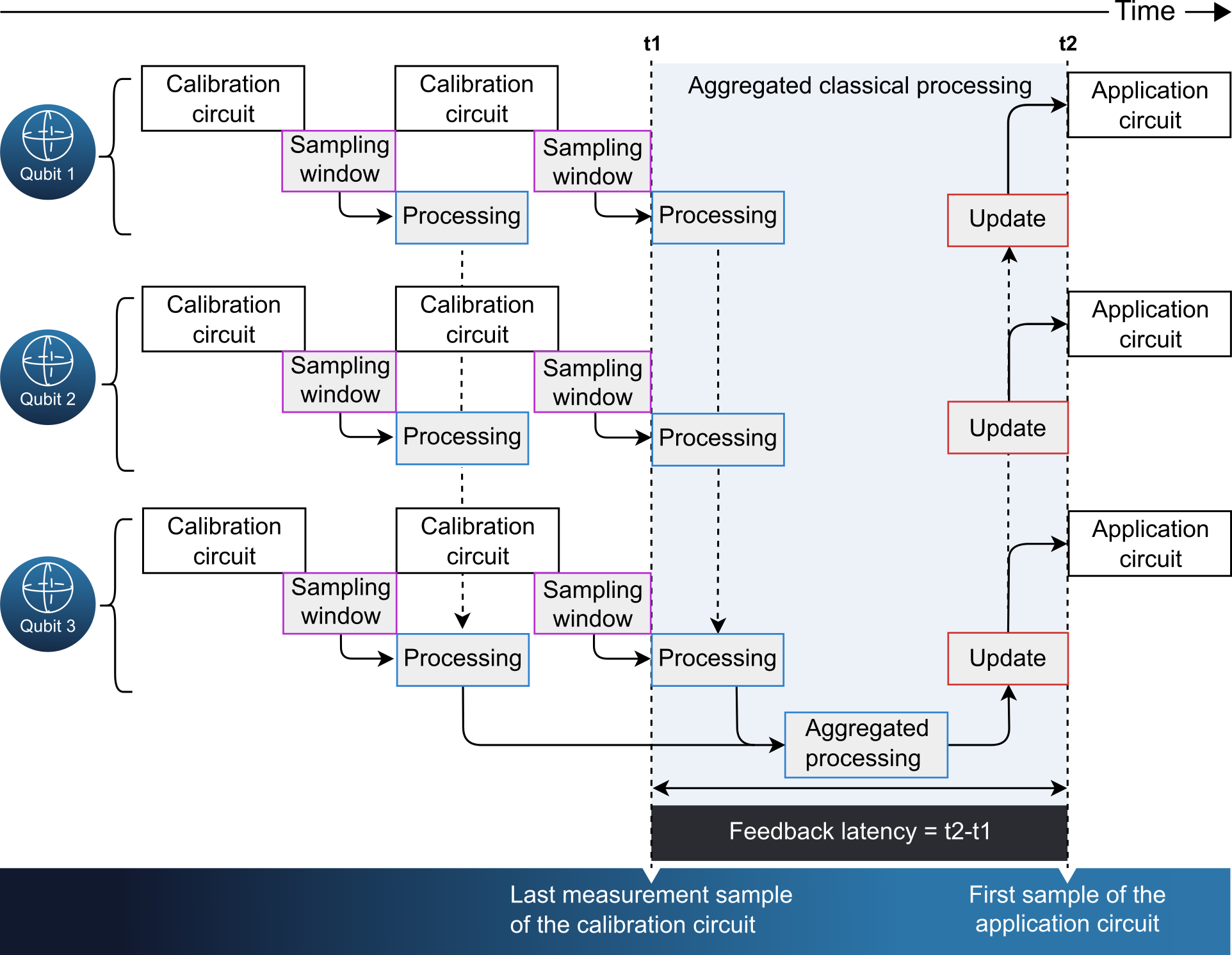}
  \caption{Timeline of controller operations for a system-real-time (SRT) quantum-classical feedback. The latency is defined as the delay from the last sample of the last measurement, representing the last calibration measurements to the first sample of the pulses updated according to the measurements statistics. The processing includes a local processing performed on each measurement and measurement round and an aggregated calculation.}
  \label{fig:fig6}
\end{figure*}

These programs include two sub-programs:

\begin{itemize}[leftmargin=*]
\item The application: the actual quantum circuit/control sequence that the user wishes to run.
\item The calibration: a sequence of quantum circuits/control sequences followed by classical processing that is applied to the measurement results, which is then used to update parameters of the application subprogram.
\end{itemize}

We define the figure of merit as the latency from the last sample of the last measurement sample (potentially of many) of the calibration subprogram to the first sample of the first control pulse of the application pulse program, as shown in Figure\,\ref{fig:fig6}.

The parameter updates for this benchmark are calculated by multiplying a dense matrix with constant entries with a vector that holds the aggregation of measurement results. The reasoning for this choice is expanded upon in Appendix\,\ref{APP2}. Below we show QUA code that defines the benchmark. Every shot, \textcolor{QM_orange}{\textbf{\textbf{N\_in}}}$\,=\,10$ measurements are performed in parallel, followed by state discrimination, and the histogram of all possible measurement results is updated. 
Note that to describe this parallelism in the code, we use a "\textbf{\textcolor{QM_purple}{for}\,\textcolor{QM_blue}{i}\,\textcolor{QM_purple}{in}\,range(\textcolor{QM_orange}{\textbf{N\_in}})}" loop notation. This does not mean that there is an actual QUA loop, but rather we use this as a compact notation to represent statements on different elements ("meta-programming"). The convention in QUA is that quantum operations on different elements (\textbf{\textcolor{QM_red}{play}}, \textbf{\textcolor{QM_red}{measure}}, \textbf{\textcolor{QM_red}{wait}}) execute in parallel.
Since updating the histogram can be parallelized with the next measurement shot, and can therefore drastically affect the benchmark results, we must add some assumption on the calibration shot duration. Here we set the calibration/measurement loop duration to be $1\,\mu s$. After \textcolor{QM_orange}{\textbf{\textbf{N\_shots}}}$\,=\,1000$ such shots, the histogram vector difference from the desired histogram ($\textbf{\textcolor{QM_red}{H0}}$ in the code) is multiplied by a constant dense matrix. The resulting vector is then used to update the parameters of the \textcolor{QM_orange}{\textbf{\textbf{N\_out}}}$\,=\,10$ subsequent pulses. Here the updated parameters are qubit control frequencies. In Appendix\,\ref{APP1} we discuss other updated parameters and variations of the benchmark.

\section{Conclusions and Outlook}  \label{CH6: Conclusions}
In this work, we identified the need to tightly integrate classical processing with quantum processing at the pulse-level. We discussed the various reasons, categorized the requirements, and presented relevant demonstrations from recent works that are relevant to propose pulse level benchmarks. We identified the quantum controller as a critical element for such integration and proposed practical tests that enable evaluation of the controller’s ability to support its requirements. The benchmarks defined in this work cover a wide range of use cases and some, namely QRT benchmarks, have been measured on Quantum Machines' control system\,\cite{WEB_feedback}, reported to be the shortest latencies in the industry. Other metrics are left for future works. In particular, benchmarks that enforce heavier classical processing, such as those required for quantum error correction and hybrid workflows involving heavy pulse optimizations, can play an important role. As the quantum computing community pushes towards quantum advantage in the NISQ era and realizing fault-tolerant quantum computing, we believe that the use of tightly integrated classical processing with ultra-low latency feedback will be critical in pushing the limits of what can be done with a given quantum hardware. We believe the above categorization and benchmarks could serve the community to move faster towards these goals. 

\section*{Acknowledgements}
We acknowledge the many contributors who provided data from their recent works involving quantum-classical processing: Will Gilbert, Andrew Dzurak, Remy Dassonneville, Ankur Agrawal and David Schuster.

\section*{Appendices}
\appendix
   
\section{Additional benchmarks and variations}  \label{APP1}

In this section, we define additional variations for the benchmarks in the main text to provide a comprehensive set covering a broader range of applications. 

\subsection*{BM1: QRT and test parametrization } 

The QRT benchmarks defined in the main text are for the case of a single controller measurement input channel and a single controller output channel. Here we expand the benchmarks to cover the multi \textit{fan-in/fan-out} case. Fan-in $(N_{in})$ is the number of controller measurement input channels, which acquire data from the quantum system (e.g., number of readout channels\cite{Devulapalli2022} in a superconducting qubit QPU), and fan-out $(N_{out})$ is the number of controller output channels (can be both control and measurement channels) affected by the measurement. Typically, the feedback latency increases as either $N_{in}$ or $N_{out}$ increase, as the communication between sub-modules of the controller adds overheads. The rate of this increase in latency as a function of these numbers is therefore an important measure of the controller. To keep things simple for this benchmark, we set $N_{inout} = N_{out} = N_{out}$ throughout. For current controller evaluations we propose to benchmark at $N_{inout} = 1,\,20,\,50$.

When dealing with the multi \textit{fan-in/fan-out} case, it is important to distinguish between two types of classical processing and feedback, which we call \textit{aggregated} and \textit{distributed} processing.  Distributed processing is where the processing and feedback process is performed in parallel (e.g., active reset performed on all qubit in parallel). Aggregated processing is where a centralized processing unit receives information from multiple channels, calculates a feedback output based on all of them, and subsequently plays to a large number of channels in parallel (e.g., in the case of a decoder in error correction protocols\cite{Das2020}). Figures\,\ref{fig:A1},\,\ref{fig:A2},\,\&\,\ref{fig:A3} demonstrate this distinction. 

In Listings\,\ref{codeA1.1.1}-\ref{codeA1.3.7} we provide the definitions of the multi \textit{fan-in/fan-out} benchmarks for each of the QRT benchmarks for both the distributed and aggregated cases (see Figures\,\ref{fig:A1},\,\ref{fig:A2}\,\&\,\ref{fig:A3}). For the conditional operations benchmarks we also add yet another variation in which the measurement results are processed before the state discrimination, which we call \textit{aggregated integer}. This is important in cases where more complex multi-channel signal processing is desired, e.g., in frequency multiplexed readout.

Regarding BM1.3 QRT parametric updates (see Figure\,\ref{fig:A3}), there are multiple parameters which can be updated in real-time, for these benchmarks, we have picked a few which have relevant use-cases: Phase/Frame (track state dependent phase, Listings\,\ref{codeA1.3.1}\,\&\,\ref{codeA1.3.2}), Frequency (track qubit frequency, Listings\,\ref{codeA1.3.3}\,\&\,\ref{codeA1.3.4}), Amplitude ($\pi-$pulse calibrations, Listings\,\ref{codeA1.3.5}\,\&\,\ref{codeA1.3.6}),  and Threshold (adaptive active reset, Listings\,\ref{codeA1.3.7}).

\subsection*{BM2: Multi-shot parameter updates modifiable parameters} 

In the main text, BM2 enforces that: 
\begin{itemize}[leftmargin=*]
\item The parameters to be updated are the frequencies of the control pulses.  
\item The matrix multiplying the histogram vector is a dense matrix 
\end{itemize}

In Table\,\ref{Tab:BM2.1param}, we list other variations of the benchmark for different parameters to be updates and different forms of the matrix, as well as note to which use cases they are relevant.

\section{Generality of the parameter update calculation}  \label{APP2}

Let us briefly discuss the problem of experimental qubit parameter estimation. A quantum circuit has an output distribution $\overrightarrow p_{ideal}$ associated with it, which is obtained in an optimally calibrated device. By repeating this circuit multiple times, we obtain an empirical outcome histogram $\overrightarrow p_{measured}$, whose entries are the measured frequencies of all the available quantum states. Reaching this optimal distribution requires knowledge of parameters describing the QPU, $\overrightarrow r_{actual}$, e.g transition frequencies, Rabi rates, coupling strengths, etc. In practice, to construct the circuit and translate it to pulses, we use an estimate of these parameters, $\overrightarrow r_{estimated}$. Thus, up to a sampling error that depends on the number of shots, the deviation between the empirical distribution and the optimal distribution is a function of our estimation error: $$\overrightarrow p_{ideal} - \overrightarrow p_{measured} = F (\overrightarrow r_{actual} - \overrightarrow r_{estimated})$$

By inverting this relation, we can update our estimate for various parameters and reduce the calibration error. 

We note that in the most general case, the complete, exponentially increasing N qubit distribution is required to obtain an estimate. However, in practice, we typically only measure subsets of the qubits that are most sensitive to the parameter we wish to estimate. For example, a 2-qubit pulse amplitude is affected, at a minimum\cite{Zajac2021}, by the two qubits participating in the gate. 

This reasoning implies that, in general, the measurement and processing flow for a parameter update takes the form shown in Figure\,\ref{fig:fig6}.  

A parameter estimate will be successful if it is close to the actual value of this parameter, but, importantly, also acquired on a timescale shorter than the parameter drift time. Therefore, we define the concept of \textit{embedded calibrations}: calibrations that run entirely from the control system and update parameters in extremely short time scales in order to both optimize performance by correcting for drifts and increase uptime by saving or reducing the need to perform more complex calibrations. Our benchmark, therefore, assumes that parameters (and by extension, the measurement outcome histogram) drift by a small amount from the optimal distribution at the time of our measurement, so we can linearize $F$ to get:  $$\overrightarrow r_{actual} = \overrightarrow r_{estimated} + M (\overrightarrow p_{ideal} - \overrightarrow p_{measured}) $$

where $$  M = \left(  \frac{\partial F}{\partial \overrightarrow r} \right)^{-1}$$

\section{Control and measurement specifications }  \label{APP3}

For the benchmarks, we define the following pulses. For control ($\pi$ and $\pi /2$) pulses, a gaussian envelope pulse is used, with sigma $4\,ns$ and played for $20\,ns$. The gaussian is modulated at $5.6\,GHz$ at the qubit's frequency. For readout pulses, a fixed amplitude pulse is used, with length of $200\,ns$, modulated at $7.6\,GHz$ at the resonator's frequency. Both pulses can either be a directly synthesized pulse at the RF frequency, or an I/Q pair with an intermediate frequency of $f_{IF} = 100\,MHz$.

For all the benchmarks in the main text, we have used the \textbf{demod} operation on the input data. The readout pulse returns to the ADC of the controller, where the \textbf{demod} operation takes the raw ADC data and process it in the following manner: $$ demod = \sum_{i=0}^{N} cos(2\pi \cdot f_{IF} \cdot t_i) S_i$$

Where $N$ is the length of the \textbf{\textcolor{QM_blue}{sampling\_window}}, and we assume that the sum starts after the $\textbf{\textcolor{QM_blue}{time\_of\_flight}}$ (see Figure\,\ref{fig:fig5}), which is when the first data point arrives to the controller. $S_i$ are the samples coming into the ADC. Note that for simplicity, we are only looking at the $I$ component of the incoming signal.

\vspace{1cm}


\def\url#1{}
\bibliographystyle{IEEEtran}  
\bibliography{IEEEabrv,QMwhitepaper}


\setcounter{figure}{0}    
\renewcommand\thefigure{A\arabic{figure}} 

\begin{figure*}
  \centering
  \includegraphics[width=2.05\columnwidth]{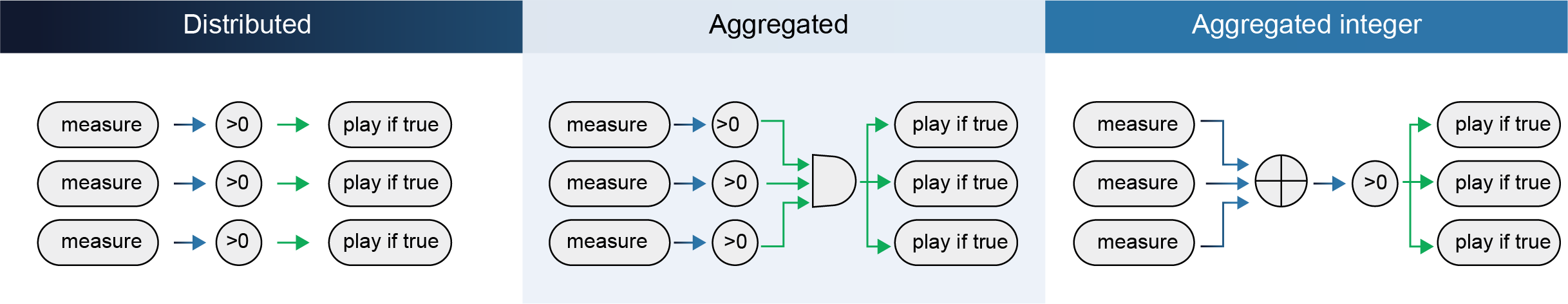}
  \caption{Different types of multi-channel QRT conditional feedback. (left) Distributed, where each measurement result independently controls a conditional play on the relevant channel. (middle) Aggregated, where each measurement is discriminated independently providing a binary value. An aggregated result is then calculated by performing an AND operation on all the binary values. The aggregated result controls a conditional play on all the channels. (right) Aggregated integer, a variation of Aggregated feedback, where the aggregated result is calculated by summing the measurements of all the channels, comparing the result to a single threshold.}
  \label{fig:A1}
\end{figure*}

\begin{figure*}
\begin{lstlisting}[language=Python, label={codeA1.1.1}, caption=QUA code outlining the BM1.1 benchmark for QRT conditional operation for the distributed case.] 
def benchmark_deterministic_QRT_conditional_distributed(N_inout): 
   int[N_inout] x 
   bool[N_inout] s_ar 
    
   strict_timing: 
      for i in range(N_inout):  # meta-programming - loop unrolled
                                # measurements are parallel
         measure(readout_pulse, readout_element[i], ...
                 demod(x[i]), timestamp -> re_time[i]) 
         s_ar[i] = x[i] > 0 
         wait(max_time=max_latency, control_element[i]) 
         play(control_pulse, control_element[i], ...
              condition=s_ar[i], timestamp -> ce_time[i]) 
                 
   return feedback_latency =max(ce_time -(re_time + time_of_flight + sampling_window)) 
\end{lstlisting}
\end{figure*}

\begin{figure*}
\begin{lstlisting}[language=Python, label={codeA1.1.2}, caption=QUA code outlining the BM1.1 benchmark for QRT conditional operation for the aggregated case.] 
def benchmark_deterministic_QRT_conditional_aggregated(N_inout): 
    int[N_inout] x 
    bool[N_inout] s_ar 
    bool s 
    
    strict_timing: 
        for i in range(N_inout):  # meta-programming - loop unrolled
                                  # measurements are parallel
            measure(readout_pulse, readout_element[i], ...
                    demod(x[i]), timestamp -> re_time[i]) 
            s_ar[i] = x[i] > 0 
        s = and(s_ar)  # calculate `and` across all elements 
        wait(max_time = max_latency, all_elements)
        align(all_elements) 
        for i in range(N_inout):  # meta-programming - loop unrolled
            play(control_pulse, control_element[i], ...
                 condition=s, timestamp -> ce_time[i] )
                 
    return feedback_latency =ce_time[0] -max(re_time +time_of_flight +sampling_window) 
    # All the control elements are aligned, thus we use the timestamp of element 0
\end{lstlisting}
\end{figure*}

\begin{figure*}
\begin{lstlisting}[language=Python, label={codeA1.1.3}, caption=QUA code outlining the BM1.1 benchmark for QRT conditional operation for the aggregated integer case.] 
def benchmark_deterministic_QRT_conditional_aggregated_int(N_inout): 
    int[N_inout] x 
    bool s 
    
    strict_timing: 
        for i in range(N_inout):  # meta-programming - loop unrolled
                                  # measurements are parallel   
            measure(readout_pulse, readout_element[i], ...
                    demod(x[i]), timestamp -> re_time[i]) 
        s = sum(x) > 0   
        wait(max_time = max_latency, all_elements)
        align(all_elements)
        for i in range(N_inout):  # meta-programming - loop unrolled 
            play(control_pulse, control_element[i], ...
                 condition=s, timestamp -> ce_time[i])
                 
    return feedback_latency =ce_time[0] -max(re_time +time_of_flight +sampling_window) 
    # All the control elements are aligned, thus we use the timestamp of element 0

\end{lstlisting}
\end{figure*}


\begin{figure*}
  \centering
  \includegraphics[width=2.05\columnwidth]{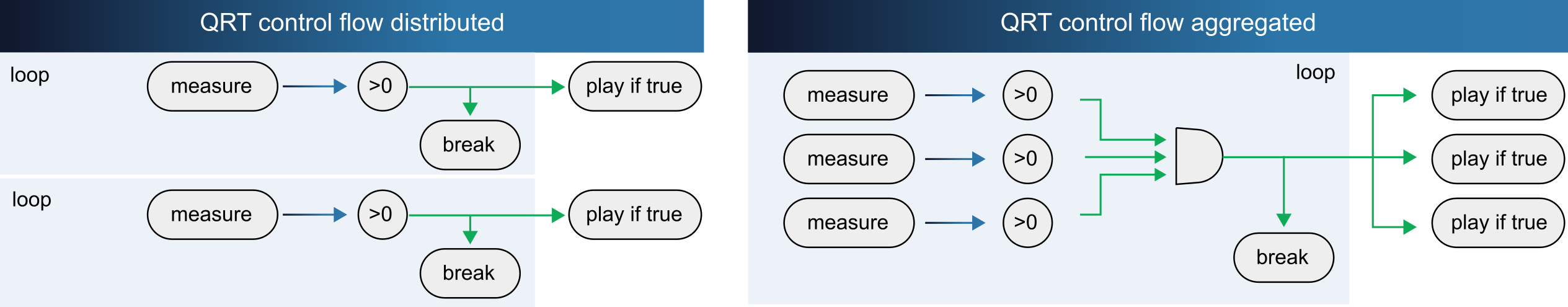}
  \caption{Different types of multi-channel QRT control flow. (left) Distributed, where each measurement result independently controls the control flow of the relevant channel. (right) Aggregated, where each measurement is discriminated independently providing a binary value. An aggregated result is then calculated by performing an AND operation on all the binary values. The aggregated result controls the control flow of all the channels.}
  \label{fig:A2}
\end{figure*}

\begin{figure*}
\begin{lstlisting}[language=Python, label={codeA1.2.1}, caption=QUA code outlining the BM1.2 benchmark for QRT control flow for the distributed case.] 
def benchmark_deterministic_QRT_control_flow_distributed(N_inout): 
    fixed[N_inout] x 
    bool[N_inout] s_ar 
    for (i, 0, i < N_inout, i+1): 
        s_ar[i] = False 
        
    strict_timing: 
        for i in range(N_inout):  # meta-programming - loop unrolled
                                  # measurements are parallel   
            while s_ar[i] == False: 
                measure(readout_pulse, readout_element[i], ...
                        demod(x[i]), timestamp -> re_time[i]) 
                s_ar[i] = x[i] > 0 
                wait(max_time=max_latency, readout_element[i]) 
            align(control_element[i], readout_element[i]) 
            play(control_pulse, control_element[i], timestamp -> ce_time[i]) 

    return feedback_latency = max(ce_time -(re_time +time_of_flight +sampling_window)) 
\end{lstlisting}
\end{figure*}

\begin{figure*}
\begin{lstlisting}[language=Python, label={codeA1.2.2}, caption=QUA code outlining the BM1.2 benchmark for QRT control flow for the aggregated case.] 
def benchmark_deterministic_QRT_control_flow_aggregated(N_inout): 
    fixed[N_inout] x 
    bool[N_inout] s_ar 
    bool s = False
    
    strict_timing: 
        while s == False:   
            for i in range(N_inout):  # meta-programming - loop unrolled
                                      # measurements are parallel   
                measure(readout_pulse, readout_element[i], ...
                        demod(x[i]), timestamp -> re_time[i]) 
                s_ar[i] = x[i] > 0 
            s = and(s_ar) # calculate `and` across all elements
            wait(max_time = max_latency, readout_elements)
        align(all_elements) 
        for i in range(N_inout):  # meta-programming - loop unrolled 
            play(control_pulse, control_element[i], timestamp -> ce_time[i]) 
            
    return feedback_latency =ce_time[0] -max(re_time +time_of_flight +sampling_window) 
    # All the control elements are aligned, thus we use the timestamp of element 0

\end{lstlisting}
\end{figure*}


\begin{figure*}
  \centering
  \includegraphics[width=2.05\columnwidth]{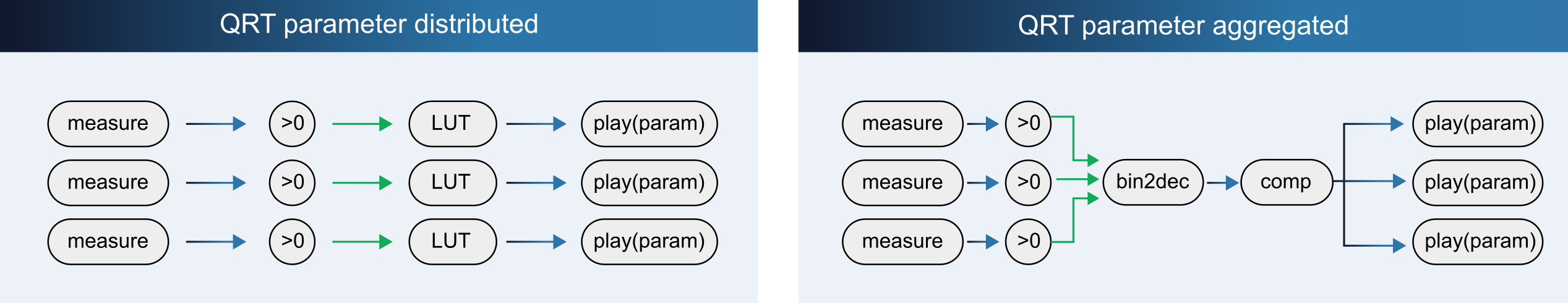}
  \caption{Different types of multi-channel QRT full parametric update feedback. (left) Distributed, where the measurement of each channel is used to query a lookup table and retrieve the updated parameter for that channel. (right) Aggregated, where each measurement is compared independently and their results are aggregated by encoding their discriminated values to a decimal value. The decimal value is then used to calculate an updated parameter for all the channels.}
  \label{fig:A3}
\end{figure*}

\begin{figure*}
\begin{lstlisting}[language=Python, label={codeA1.3.1}, caption=QUA code outlining the BM1.3 benchmark for QRT parametric updates of phase/frame (distributed).] 
def benchmark_deterministic_QRT_parametric_frame_distributed(N_inout): 
    fixed[N_inout] x 
    fixed[N_inout] frame_rot_ang_ar 
    fixed[2] frame_lut = [0.1, 0.2] 
    
    strict_timing: 
        for i in range(N_inout):  # meta-programming - loop unrolled
                                  # measurements are parallel  
            measure(readout_pulse, readout_element[i], ...
                    demod(x[i]), timestamp -> re_time[i]) 
            frame_rot_ang_ar[i] = frame_lut[x[i] > 0] 
            wait(max_time=max_latency, control_element[i]) 
            frame_rotation_2pi(frame_rot_ang_ar[i], control_element[i]) 
            play(control_pulse, control_element[i], timestamp -> ce_time[i])  
            
    return feedback_latency = max(ce_time -(re_time +time_of_flight +sampling_window)) 
\end{lstlisting}
\end{figure*}

\begin{figure*}
\begin{lstlisting}[language=Python, label={codeA1.3.2}, caption=QUA code outlining the BM1.3 benchmark for QRT parametric updates of phase/frame (aggregated).] 
def benchmark_deterministic_QRT_parametric_frame_aggregated(N_inout):  
    fixed[N_inout] x 
    bool[N_inout] s_ar 
    fixed frame_rot_ang 
 
    strict_timing: 
        for i in range(N_inout):  # meta-programming - loop unrolled
                                  # measurements are parallel 
            measure(readout_pulse, readout_element[i], ...
                    demod(x[i]), timestamp -> re_time[i]) 
            s_ar[i] = x[i] > 0 
        frame_rot_ang = bin2dec(s_ar) / 2 ** N_inout  
        align(all_elements)             
        wait(max_time = max_latency, control_elements)
        for i in range(N_inout):  # meta-programming - loop unrolled
            frame_rotation_2pi(frame_rot_ang, control_element[i]) 
            play(control_pulse, control_element[i], timestamp -> ce_time[i]) 
 
    return feedback_latency =ce_time[0] -max(re_time +time_of_flight +sampling_window) 
    # All the control elements are aligned, thus we use the timestamp of element 0

\end{lstlisting}
\end{figure*}

\begin{figure*}
\begin{lstlisting}[language=Python, label={codeA1.3.3}, caption=QUA code outlining the BM1.3 benchmark for QRT parametric updates of frequency (distributed).] 
def benchmark_deterministic_QRT_parametric_frequency_distributed(N_inout):  
    fixed[N_inout] x 
    int[N_inout] frequency_ar 
    int[2] frequency_lut = [50e6, 70e6] 
 
    strict_timing: 
        for i in range(N_inout):  # meta-programming - loop unrolled
                                  # measurements are parallel 
            measure(readout_pulse, readout_element[i], ...
                    demod(x[i]), timestamp -> re_time[i]) 
            frequency_ar[i] = frame_lut[x[i] > 0] 
            wait(max_time=max_latency, control_element[i]) 
            update_frequency(frequency_ar[i], control_element[i]) 
            play(control_pulse, control_element[i], timestamp -> ce_time[i])  
            
    return feedback_latency = max(ce_time -(re_time +time_of_flight +sampling_window)) 
\end{lstlisting}
\end{figure*}

\begin{figure*}
\begin{lstlisting}[language=Python, label={codeA1.3.4}, caption=QUA code outlining the BM1.3 benchmark for QRT parametric updates of frequency (aggregated).] 
def benchmark_deterministic_QRT_parametric_frequency_aggregated(N_inout): 
    fixed[N_inout] x 
    bool[N_inout] s_ar 
    int frequency_update  
    
    strict_timing: 
        for i in range(N_inout):  # meta-programming - loop unrolled
                                  # measurements are parallel 
            measure(readout_pulse, readout_element[i], ...
                    demod(x[i]), timestamp -> re_time[i]) 
            s_ar[i] = x[i] > 0 
        frequency_update = 100e6 * bin2dec(s_ar) / 2 ** N_inout  
        align(all_elements)
        wait(max_time = max_latency, control_elements)
        for i in range(N_inout):  # meta-programming - loop unrolled
            update_frequency(frequency_update, control_element[i])
        align(control_elements)
        for i in range(N_inout):  # meta-programming - loop unrolled
            play(control_pulse, control_element[i], timestamp -> ce_time[i])  
             
    return feedback_latency =ce_time[0] -max(re_time +time_of_flight +sampling_window)
    # All the control elements are aligned, thus we use the timestamp of element 0

\end{lstlisting}
\end{figure*}

\begin{figure*}
\begin{lstlisting}[language=Python, label={codeA1.3.5}, caption=QUA code outlining the BM1.3 benchmark for QRT parametric updates of amplitude (distributed).] 
def benchmark_deterministic_QRT_parametric_amp_distributed(N_inout): 
    fixed[N_inout] x 
    fixed[N_inout] amp_ar 
    fixed[2] amp_lut = [0.7, 0.9]

    strict_timing: 
        for i in range(N_inout):  # meta-programming - loop unrolled
                                  # measurements are parallel  
            measure(readout_pulse, readout_element[i], ...
                    demod(x[i]), timestamp -> re_time[i]) 
            amp_ar[i] = amp_lut[x[i] > 0] 
            wait(max_time=max_latency, control_element[i]) 
            play(control_pulse * amp(amp_ar[i]), ...
                 control_element[i], timestamp -> ce_time[i]) 
            
    return feedback_latency = max(ce_time -(re_time +time_of_flight +sampling_window)) 
\end{lstlisting}
\end{figure*}

\begin{figure*}
\begin{lstlisting}[language=Python, label={codeA1.3.6}, caption=QUA code outlining the BM1.3 benchmark for QRT parametric updates of amplitude (aggregated).] 
def benchmark_deterministic_QRT_parametric_amp_aggregated(N_inout): 
    fixed[N_inout] x 
    bool[N_inout] s_ar 
    fixed amp_update 

    strict_timing: 
        for i in range(N_inout):  # meta-programming - loop unrolled
                                  # measurements are parallel 
            measure(readout_pulse, readout_element[i], ...
                    demod(x[i]), timestamp -> re_time[i]) 
            s_ar[i] = x[i] > 0 
        amp_update = bin2dec(s_ar) / 2 ** N_inout 
        align(all_elements) 
        wait(max_time = max_latency, control_elements)
        for i in range(N_inout):  # meta-programming - loop unrolled
            play(control_pulse * amp(amp_update), ...
                 control_element[i], timestamp -> ce_time[i]) 
            
    return feedback_latency =ce_time[0] -max(re_time +time_of_flight +sampling_window)
    # All the control elements are aligned, thus we use the timestamp of element 0

\end{lstlisting}
\end{figure*}

\begin{figure*}
\begin{lstlisting}[language=Python, label={codeA1.3.7}, caption=QUA code outlining the BM1.3 benchmark for QRT parametric updates of threshold (distributed).] 
def benchmark_deterministic_QRT_parametric_threshold_distributed(N_inout): 
    fixed[N_inout] x1 
    fixed[N_inout] x2 
    fixed[N_inout] threshold_ar 
    fixed[2] threshold_lut = [0.1, 0.2] 
    bool[N_inout] s_ar 

    strict_timing: 
        for i in range(N_inout):  # meta-programming - loop unrolled
                                  # measurements are parallel 
            measure(readout_pulse, readout_element[i], ...
                    demod(x1[i]), timestamp -> re_time[i]) 
            threshold_ar[i] = threshold_lut[x1[i] > 0] 
            measure(readout_pulse, readout_element[i], demod(x2[i])) 
            s_ar[i] = x2[i] > threshold_ar[i] 
            wait(max_time=max_latency, control_element[i]) 
            play(control_pulse, control_element[i], ...
                 condition= s_ar[i], timestamp -> ce_time[i]) 
            
    return feedback_latency = max(ce_time -(re_time +time_of_flight +sampling_window)) 
\end{lstlisting}
\end{figure*}


\setcounter{figure}{0}    

\renewcommand{\figurename}{Table}
\begin{figure*}
\begin{tabular}{|p{0.21\linewidth}|p{0.15\linewidth}|p{0.55\linewidth}|} 
\hline
\textbf{Parameter to update}       & \textbf{Matrix form}            & \textbf{Use case}                                                                                                                                       \\
\hline
Control pulse frequency            & diagonal                        & Frequency re-calibration via simultaneous Ramsey on all qubits at given delay time.                                                                     \\
\hline
\multirow{2}{*}{Control pulse amplitude} & \multirow{2}{*}{diagonal}             & XY drive amplitude re-calibrations via simultaneous pi/2 pulse on all qubits.                                                                           \\
\cline{3-3}
                                   &                                 & Fast flux/gate pulse re-calibrations in SC/Spin qubits via simultaneous Ramsey on all qubits at given delay time.                                       \\
\hline
Control pulse amplitude            & $2x2$ block diagonal            & Fast flux/gate pulse re-calibrations in SC/Spin qubits via simultaneous two qubit gates between a pair of qubits which maximally cover the qubits grid  \\
\hline
DC offset                          & Full (all entries are non-zero) & Flux/gate working point re-calibrations in SC/Spin qubits via simultaneous Ramsey on all qubits at given delay time                                     \\
\hline
Discrimination threshold           & Full (all entries are non-zero) & Discrimination threshold re-calibrations via simultaneous pi/2 pulse on all qubits                                                                      \\
\hline

\end{tabular}
\caption{\label{Tab:BM2.1param}Table listing other variants of the benchmark BM2 for multi-shot parameter updates, for different parameters to be updated and different forms of the matrix, as well as note to which use case they are relevant.}
\end{figure*}


\end{multicols}
\end{document}